\documentclass[journal]{IEEEtran}
\usepackage{amssymb,amsmath,theorem}
\usepackage{array}
\usepackage{graphicx}
\usepackage{epsfig}
\usepackage{amsfonts}
\usepackage{bm}
\usepackage{arydshln}
\usepackage{subfig}
\usepackage{algorithm}
\usepackage{algorithmic}
\usepackage{stfloats}

\newcolumntype{C}[1]{>{\centering}p{#1}}
\usepackage{rotating,setspace,latexsym,amsmath,epsf,amssymb,bm,color}
\usepackage{cite}
\usepackage{booktabs}

\IEEEoverridecommandlockouts

\ifCLASSINFOpdf
\else
\fi

\begin{document}
%
\title{Experimental Analysis of Multipath Characteristics in Indoor Distributed Massive MIMO Channels}

%

\author{Yingjie~Xu,~\IEEEmembership{Student member,~IEEE},
        Xuesong~Cai,~\IEEEmembership{Senior member,~IEEE},
        Sara~Willhammar,~\IEEEmembership{Member,~IEEE}, and~Fredrik~Tufvesson,~\IEEEmembership{Fellow,~IEEE}
\thanks{This work was supported by the Swedish strategic research area ELLIIT, by NextG2Com funded by the VINNOVA program for Advanced Digitalisation with grant number 2023-00541, and by REINDEER which has received funding from the European Union’s Horizon 2020 research and innovation program under grant agreement No. 101013425. \emph{(Corresponding author: Yingjie Xu.)}

Y.~Xu, S.~Willhammar, and F.~Tufvesson are with the Department of Electrical and Information Technology, Lund University, Lund, Sweden (e-mail: \{yingjie.xu, sara.willhammar, fredrik.tufvesson\}@eit.lth.se).

X. Cai is with the School of Electronics, Peking University,
Beijing, 100871, China (email: xuesong.cai@pku.edu.cn).}}

%
%

\markboth{Journal of \LaTeX\ Class Files,~Vol.~xx, No.~x, December~2024}%
{Shell \MakeLowercase{\textit{et al.}}: Bare Demo of IEEEtran.cls for IEEE Journals}
%



\maketitle

\begin{abstract}
Distributed multiple-input multiple-output (MIMO), also known as cell-free massive MIMO, has emerged as a promising technology for sixth-generation (6G) wireless networks. 
This letter introduces an indoor channel measurement campaign designed to explore the behavior of multipath components (MPCs) in distributed MIMO channels. Fully coherent channels were measured between eight distributed uniform planar arrays (128 elements in total) and a 12-meter user equipment route. A method is introduced to determine the order (single- or multi-bounce) of MPC interaction by leveraging map information and MPC parameters. In addition, a Kalman filter-based framework is used for identifying the MPC interaction mechanisms (reflection or scattering/diffraction/mixed). Finally, a comprehensive MPC-level characterization is performed based on the measured channels, including the significance of single-bounce MPCs, common single-bounce MPCs, spherical wavefront features, birth-and-death processes of the MPCs, and spatial distribution of reflections. The findings serve as a valuable reference for understanding the propagation behavior of MPCs, which is necessary for future modeling of indoor distributed MIMO channels.    
\end{abstract}

%
\IEEEpeerreviewmaketitle

\section{Introduction}
\IEEEPARstart{A}{s} a promising technology for sixth-generation (6G) wireless networks, distributed multiple-input multiple-output (MIMO), also known as cell-free massive MIMO~\cite{H. Q. Ngo2017}, is expected to combine the advantages of both massive MIMO and cell-free networks. Compared to co-located massive MIMO, distributed MIMO aims to support more uniform coverage and improve communication reliability~\cite{H. A. Ammar 2022}.  

In order to optimize the design of distributed MIMO systems, it is necessary to perform measurement campaigns to experimentally investigate their underlying propagation channels. Extensive research has been carried out in this regard. For example, outdoor channel measurements were reported in~\cite{D. Lschenbrand 2022} with a widely distributed antenna array spread over a 46.5~m range. Key channel properties, including time-variant delay spread, Doppler spread, and path loss, were investigated. Channel measurements in~\cite{Y. Zhang2024} analyzed multi-link channels between 80 user equipment (UE) and more than 20,000 access point~(AP) locations in terms of path loss, shadowing, and delay spread. The distributed MIMO channels in an industrial office were investigated in~\cite{T. Choi2020} from a perspective of multi-user capacity performance through various antenna topologies. In addition, a system with twelve fully coherent distributed antennas was introduced in~\cite{C. Nelson2025} for industrial channel measurements. Further measurements, as reported in our previous work~\cite{Y. Xu202403,Y. Xu202410}, investigate user separability and non-wide-sense stationarity in measured channels.

The aforementioned studies are limited in several aspects. First, measurements in~\cite{Y. Zhang2024,C. Nelson2025} focused solely on channels with fully distributed topologies, without considering semi distributed configurations. However, the study in~\cite{T. Choi2020} has found that semi-distributed APs can achieve improved spectral efficiency, making them a promising and practical solution for real-world deployments. Second, the aforementioned studies provide limited small-scale fading analysis, particularly multipath effects, which are critical to accurately understanding and optimizing system performance.
Studies in~\cite{W. Fan2024,Z. Yuan2024,P. Tang2024} found that the deployment of massive antennas results in spatially varying multipath components (MPCs). 
This is even more evident in distributed MIMO due to the larger separation between APs. 
Understanding MPC-level channel characterization is not only fundamental for accurately parameterization of ray-based channel models, but is also important in system design and practical applications. 

Before proposing a comprehensive parametric channel model for (semi) distributed MIMO, it is necessary to perform real-world measurements to fully explore the MPC-level characteristics in the associated channels. However, to the best of our knowledge, this has not been completely addressed. To fill this gap, this letter presents an indoor distributed MIMO channel measurement campaign. Then, two methods are introduced to determine the order of the MPC interaction and their interaction mechanisms. Finally, MPC characteristics of the measured channels are investigated, including the significance of single-bounce MPCs, the spherical wavefront properties, as well as birth-and-death processes of the MPCs, and the spatial distribution of reflections within the channels. The findings provide a detailed understanding of the characteristics at the MPC level of distributed MIMO channels, which form the basis for future channel modeling.  

\section{Indoor Distributed MIMO Channel Measurements}
The indoor channel measurements were performed in a laboratory room at Lund University, Sweden. The room features an environment with various objects that contribute to a rich scattering environment. The signals can interact with these objects through reflections, diffraction, or scattering, allowing for MPC investigations with diverse propagation behavior.

\begin{figure}[tb]
	\centering
	{
		\begin{minipage}[tb]{0.50\textwidth}
			\centering
			\includegraphics[width=1\textwidth]{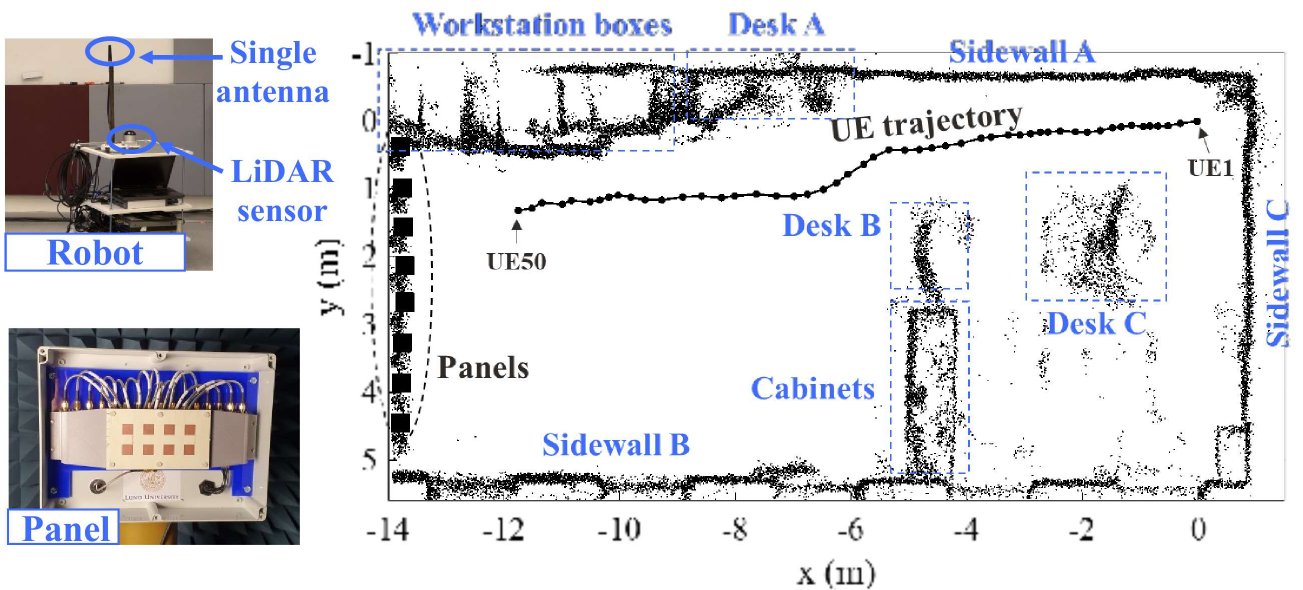}
		\end{minipage}
	}
	\caption{UE-side antenna, BS-side panel, and Lidar point cloud of the measurement environment.}
	\label{Fig_Measurement_environment}
\end{figure}

A switch-based wideband distributed MIMO channel sounder~\cite{M. Sandra2024} is deployed, integrating multiple NI universal software radio peripherals (USRPs), SP16T radio frequency (RF) switches, and a pair of Rubidium clocks. Eight uniform planar arrays (UPAs) are deployed on the base station (BS) side, here referred to as `panels'. As illustrated in Fig.~\ref{Fig_Measurement_environment}, each panel is equipped with eight dual-polarized patch antennas (16 ports), designed for a 5--6~GHz frequency band (S11$<-10$~dB). The spacing between neighboring patch antennas is 26.7~mm, corresponding to approximately half a wavelength in this frequency range. On the UE side, a single omnidirectional monopole antenna was used with an operating frequency band of 3.5--8~GHz. Note that despite the single-antenna configuration at the UE side, the overall system still qualifies as a distributed MIMO channel measurement setup, particularly from the perspective of the BS, which transmits and receives over multiple spatially separated antenna elements. This configuration is consistent with widely adopted practices in MIMO channel measurements~\cite{X. Gao2013,X. Gao2015}.

The measurements were performed at a carrier frequency of 5.7~GHz with a bandwidth of 400~MHz. In the measurement, the UE antenna was mounted on a robot, as shown in Fig.~\ref{Fig_Measurement_environment}, which moved from one end of the room toward the panels with a speed of 0.012~m/s. The robot's trajectory was recorded by a Lidar sensor, which simultaneously collected a point cloud of the environment, as shown in Fig.~\ref{Fig_Measurement_environment}. The panels were distributed on one side of the room, with a spacing of 60~cm between each other. This setup can be viewed as a semi-distributed AP topology. In total, uplink channels were measured along a 12-meter UE route and channel impulse responses (CIRs) were collected from 50 different positions (snapshots). To extract MPC parameters from the CIRs, including the complex polarization matrix, angle of arrival (AoA), delay, and Doppler frequency, the space-alternating generalized expectation maximization (SAGE) algorithm~\cite{X. Yin 2003} was applied. 

\section{MPC Propagation Characteristics}
\subsection{MPC interaction order (single-bounce vs. multi-bounce)}
During propagation, interactions with physical objects may occur along each MPC. Based on the number of interactions that occurred, MPCs can be categorized as single-bounce (only a single interaction) or multi-bounce (multiple interactions). Distinguishing the interaction order of each MPC is essential for channel characterization, modeling, and environmental mapping. 
In this section, a method for distinguishing MPCs with different interaction orders is introduced, harnessing the extracted MPC parameters and the Lidar point cloud.

Considering a double-directional channel model, each MPC is associated with a first-hop and a last-hop scatterer/interaction. For single-bounce MPCs, their associated interactions coincide. Since only AoAs of the MPCs are available in our measurements, the position of the last-hop scatters is calculated. Let $\mathbf{s}_{l,k}^\text{LH}=(x_{s},y_{s},z_{s})$ denote the position of the last-hop scatter of MPC~${l}$ in the panel~$k$-UE channel link. Given the angle of arrival $\mathbf{\Omega}_{l,k} =\left [ \phi_{l,k}, \theta_{l,k}  \right ]$, define a ray $\xi_{l,k}$ starting from the location of panel~$k$, i.e., $\mathbf{P}_k=(x_{p_k},y_{p_k},z_{p_k})$ with the parametric equation of
\begin{equation}
\xi_{l,k}: \begin{cases}
x = x_{p_k}+d \cdot \cos \phi_{l,k} \cdot \sin \theta_{l,k}, \\
y = y_{p_k}+d \cdot \sin \phi_{l,k} \cdot \sin \theta_{l,k}, \\
z = z_{p_k}+d \cdot \cos \theta_{l,k},
\end{cases}
\label{MPC_ray}
\end{equation}
where $0< d\leq \tau_{l,k}\cdot c$ is a distance parameter, where $\tau_{l,k}$ is the delay of MPC~${l}$ and $c$ represents the speed of light. Then $\mathbf{s}_{l,k}^\text{LH}=(x_{s},y_{s},z_{s})$ can be obtained by finding the first valid intersection of $\xi_{l,k}$ and the point cloud, as described in~\cite{Y. Xu2025}.

For clarity, let hypothesis $\zeta=\zeta_1$ represent that MPC~$l$ is single-bounce, while the alternate hypothesis $\zeta=\zeta_2$ indicates that it is multi-bounce. Given the condition $\zeta=\zeta_1$, the position of the associated scatterer of MPC~$l$ is estimated by solving an optimization problem as 
\begin{equation}
\hat{\mathbf{s}}_{l,k}=\min_{\mathbf{s}_{l,k}}\left \| \mathbf{s}_{l,k}-\mathbf{P}_k-(\tau_{l,k}\cdot c-\left \|\mathbf{s}_{l,k}-\mathbf{u} \right \|_\text{F} )\cdot \mathbf{e}_{l,k} \right \|_\text{F}^2,  
\label{}
\end{equation}
where $\mathbf{u}$ and $\left \|\cdot \right \| _\text{F}$ represent the UE position and the Frobenius norm operation, respectively, and the norm vector $\mathbf{e}_{l,k}=[\cos \phi_{l,k}\sin \theta_{l,k}, \sin \phi_{l,k}\sin \theta_{l,k}, \cos \theta_{l,k}]^\text{T}$. Theoretically, whether MPC~${l}$ is single-bounce or not can be determined by checking if $\hat{\mathbf{s}}_{l,k}=\mathbf{s}_{l,k}^\text{LH}$. Taking into account the potential measurement noise and estimation uncertainty, the interaction order of MPC~${l}$ is determined by a metric as
\begin{equation}
 \zeta=\begin{cases}
\zeta_1,\quad \left \|\hat{\mathbf{s}}_{l,k}-\mathbf{s}_{l,k}^\text{LH} \right \|_\text{F}\leq \delta_\zeta,  \\
\zeta_2,\quad\text{others},
\end{cases}
\label{}
\end{equation}
where $\delta_\zeta$ is a preset threshold. Note that $\delta_\zeta$ should be large enough to accept MPCs that follow the single-bounce hypothesis, while remaining small enough to reject those that involve multiple bounces, but may otherwise appear to match the single-bounce condition. A sigma-point method, as described in~\cite{I. Arasaratnam2009}, is employed here to choose an appropriate $\delta_\zeta$. Due to space limitations, details are omitted. Based on the evaluation, the optimal $\delta_\zeta$ is set to 1.5 for the subsequent analysis.
\subsection{MPC interaction mechanisms (reflection vs. others)}
\begin{figure}[tb]
	\centerline{\includegraphics[width=0.40\textwidth]{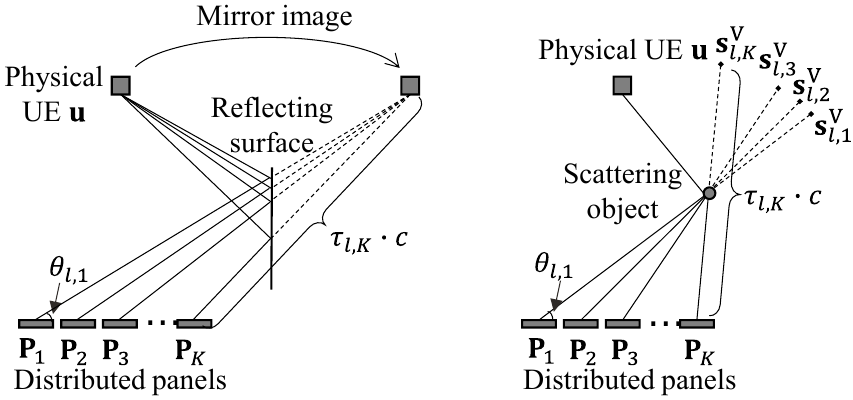}}
	\caption{Virtual scatterers with different MPC interaction mechanisms: single-bounce reflection (left) and single-bounce scattering (right).}
	\label{Fig_propagation mechanisms}
\end{figure}
Due to the complex scattering environment, MPCs may interact with physical objects through various mechanisms, including reflection, diffraction, scattering, or combinations thereof. Investigating the interaction mechanisms of MPCs is necessary to understand the spatial consistency of the channel. In particular, MPCs experiencing only specular reflections, here referred to as `reflected MPCs', have attracted significant research interest. Since these MPCs are deterministically related to both the geometry of the environment and the transceiver, they offer greater potential for various applications, such as beamforming tracking, user localization, and environmental sensing.  

To identify the reflected MPCs, first, given an MPC~$l$ from the panel~$k$-UE link with the delay $\tau_{l,k}$ and AoA $\mathbf{\Omega}_{k,l} =\left [ \phi_{k,l}, \theta_{k,l}  \right ]$, define its virtual scatterer~(VS) as $\mathbf{s}_{l,k}^\text{v}=(x_{s}^\text{v},y_{s}^\text{v},z_{s}^\text{v})=( x_{p_k}+\tau_{l,k}c\cdot \cos \phi_{k,l}  \sin \theta_{k,l},\quad y_{p_k}+\tau_{l,k}c \cdot \sin \phi_{k,l}  \sin \theta_{k,l},\quad z_{p_k}+\tau_{l,k}c \cdot \cos \theta_{k,l})$.  
For the same single-bounce reflected MPCs from different panel-UE channels, their VSs coincide with the mirror image of the UE, as shown in Fig.~\ref{Fig_propagation mechanisms}. In the case of multi-bounce reflected MPCs, their VSs will be the iterated mirror image of the UE. In contrast, for MPCs from different links but involving other interaction mechanisms, their VSs occupy distinct positions. Inspired by these features, a Kalman filter-based framework is used to track coincident VSs from different links and subsequently identify their associated MPCs as reflected MPCs. Specifically, assuming that the VSs are static within one snapshot, the state model of $\mathbf{s}_{l,k}^\text{v}$ is given by $\mathbf{s}_{l,k}^\text{v}=\mathbf{s}_{l,k-1}^\text{v}+\mathbf{w}_k$, where $\mathbf{w}_k$ is denoted as the state processing noise with covariance matrix $\mathbf{Q}$. The measurement/observation model is expressed as $\mathbf{z}_{l,k}^\text{v}=\mathbf{s}_{l,k-1}^\text{v}+\mathbf{v}_k$, where $\mathbf{z}_{l,k}^\text{v}$ and $\mathbf{v}_k$ are the locations of the measured VS and the measurement noise with the covariance matrix $\mathbf{R}$, respectively. Given $\mathbf{s}_{l,k-1}^\text{v}$ from the panel~($k-1$)-UE link, the predicted VS $\mathbf{s}_{l,k|k-1}^\text{v}$ and the error covariance matrix $\mathbf{M}_{k|k-1}$ at panel~$k$-UE link are obtained by $\mathbf{s}_{l,k|k-1}^\text{v}=\mathbf{s}_{l,k-1}^\text{v}$ and $\mathbf{M}_{k|k-1}=\mathbf{M}_{k-1}+\mathbf{Q}$, respectively. Then, data association between the predicted $\mathbf{s}_{l,k|k-1}^\text{v}$ and the measured $\mathbf{z}_{l,k}^\text{v}$ is performed. The mutual minimum distance method~\cite{X. Cai2020} is exploited here. Specifically, the distance between any $\mathbf{s}_{l,k|k-1}^\text{v}$ and $\mathbf{z}_{j,k}^\text{v}$ is calculated as $d(i,j)=\left \|\mathbf{s}_{l,k|k-1}^\text{v}-\mathbf{z}_{j,k}  \right \|_\text{F}$. Only if a predicted VS and a measured VS are mutually closest, they are tracked as the same VS. After determining the observed $\mathbf{z}_{l,k}$ for $\mathbf{s}_{l,k|k-1}^\text{v}$, the Kalman update procedure is performed as
$K_{k} = \mathbf{M}_{k|k-1} \left( \mathbf{M}_{k|k-1}  + \mathbf{R} \right)^{-1}$, $\mathbf{M}_{k} = \left( \mathbf{I} - K_{k} \right) \mathbf{M}_{k|k-1}$, and $\mathbf{s}_{l,k}^\text{v} = \mathbf{s}_{l,k|k-1}^\text{v} + K_k \left( \mathbf{z}_{l,k}- \mathbf{s}_{l,k|k-1}^\text{v} \right)$,
where $K_{k}$, $\mathbf{M}_{k}$, $\mathbf{s}_{l,k}^\text{v}$ represent the Kalman gain, the updated error covariance matrix, and the updated VS from the panel~$k$-UE link, respectively. Finally, the reflected MPCs are determined by finding whose VSs are tracked based on the above framework. At the same time, the location of the VS is estimated iteratively.  
\section{Results and Analysis}
In this section, the characteristics of MPCs in measured channels are investigated in terms of their interaction order and interaction mechanisms. 

\begin{figure}[t]
\centerline{\includegraphics[width=0.49\textwidth]{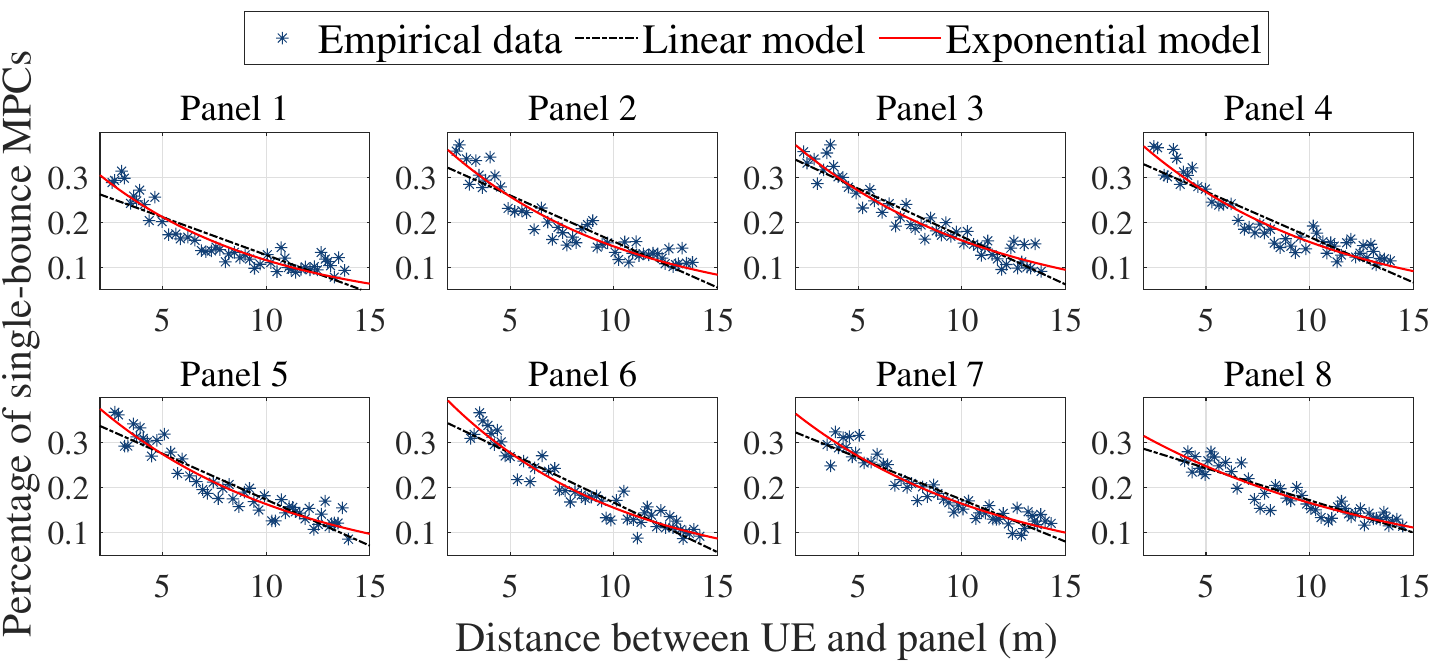}}
	\caption{Percentage of the single-bounce MPCs vs. communication distance.}
	\label{Fig_ratio_of_SB_MPC}
\end{figure}

The percentage $\eta_{\text{SB}}$ of single-bounce MPCs in different panel-UE links is shown in Fig.~\ref{Fig_ratio_of_SB_MPC}. The results in all links exhibit a consistent trend: The percentage decreases with increasing communication distance $d_c$ between the UE and the panels. It is reasonable since a longer communication distance makes an MPC less likely to reach the receiver through a single interaction, i.e. the possibility of experiencing multiple interactions is higher. Moreover, during the movement of the UE, the measured $\eta_{\text{SB}}$ of all links is below 0.4. This indicates that a significant part of the MPCs in the channels are multi-bounce, exhibiting the complex scattering environment of the measured scenario. The relationship between the measured $\eta_{\text{SB}}$ and $d_c$ is further fitted with linear and exponential models. The resulting fitting parameters and the R-squared ($R^2$) are summarized in Table~\ref{Table_fitting parameter1}. From the results for the same link, the exponential model yields a larger $R^2$, suggesting that the exponential model is preferable to characterize the relationship between $\eta_{\text{SB}}$ and $d_c$ in the measured channels. It should be noted that more measurements may be needed, particularly at larger distances, to further validate this trend.   
\begin{table}[t]
    \footnotesize
\centering
    \caption{Statistical model for the percentage of single-bounce MPCs}

    \label{Table_fitting parameter1}
    
    \begin{tabular}{c|c|c|c|c|c|c} 
    \hline
    \hline
         &\multicolumn{3}{c|}{Linear model}&\multicolumn{3}{c}{Exponential model}  \\
        &  \multicolumn{3}{c|}{$\eta_{\text{SB}}=a\cdot d_c+b$}&  \multicolumn{3}{c}{$\eta_{\text{SB}}=a\cdot e^{b\cdot d_c}$}  \\
        \hline
         Panel&  $a$&$b$&$R^2$&$a$&$b$&$R^2$  \\
\hline
         1& -0.017 &0.296  &0.786  &0.388  &-0.121  &0.883 \\
\hline
         2&  -0.017&  0.362&  0.854& 0.452&  -0.113&  0.914 \\
\hline
         3&  -0.021&  0.381&  0.9& 0.459&  -0.106&  0.924\\
\hline
         4&  -0.020&   0.369& 0.851& 0.458&  -0.101&  0.914\\
\hline
         5&  -0.021&  0.379&  0.876&  0.456& -0.108&  0.924\\
\hline
         6&  -0.022&  0.389&  0.895&  0.463 &  -0.104&  0.914\\
\hline
         7&   -0.019&  0.360& 0.884& 0.497 &  -0.116&  0.916\\
\hline
         8&  -0.014&  0.315&  0.840&  0.445&  -0.099&  0.925\\
\hline
\hline
    \end{tabular}
    
\end{table}
\begin{figure}[t]
\centerline{\includegraphics[width=0.40\textwidth]{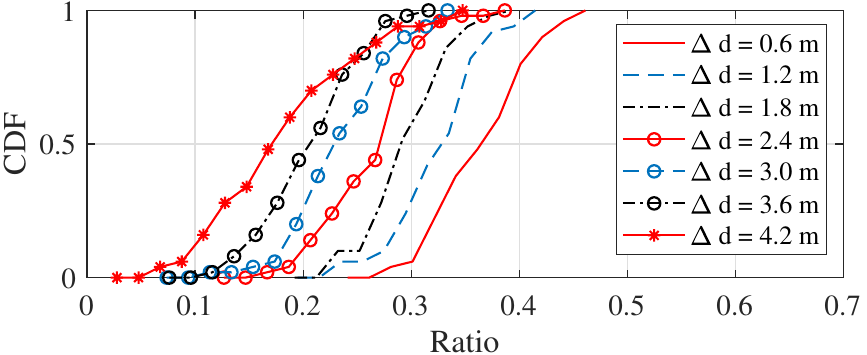}}
	\caption{CDFs of common single-bounce MPC ratios between the panels with different spacing $\Delta d$.}
	\label{Fig_cdf_common_SBMPC_raio}
\end{figure}
\begin{figure}[t]
\centerline{\includegraphics[width=0.42\textwidth]{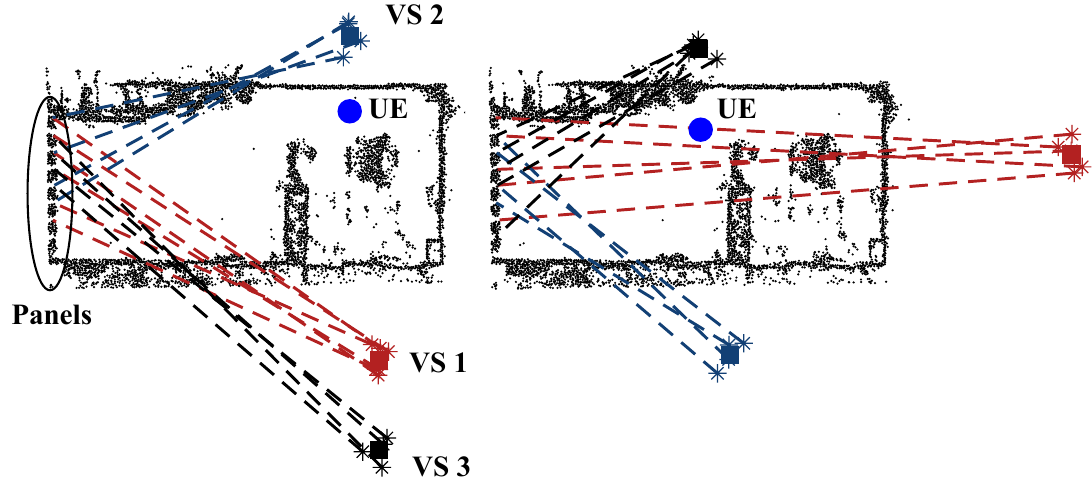}}
	\caption{Example of the reflected MPCs at snapshot~15~(left) and snapshot 29~(right), visualized on the captured Lidar point cloud.}
	\label{Fig_example_reflected_mpc_sn14sn29}
\end{figure}

To further evaluate the single-bounce MPCs among the panels, we investigate the ratios of common single-bounce MPCs. Common single-bounce MPCs are defined as the same MPCs observed from different panels. These can be identified by examining whether their virtual scatterers (for reflected MPCs) or interacting scatterers (for scattered MPCs) coincide across panels, as illustrated in Fig.~\ref{Fig_propagation mechanisms}. The cumulative distribution functions (CDFs) of the common single-bounce MPC ratios for different panel spacings $\Delta d$ are presented in Fig.~\ref{Fig_cdf_common_SBMPC_raio}. It can be seen that the ratios decrease as the panel spacing increases. This trend can be explained by the fact that as the distance between panels increases, the associated channels exhibit more diverse propagation characteristics, leading to fewer common MPCs.

For the MPC interaction mechanisms in the measured channels, Fig.~\ref{Fig_example_reflected_mpc_sn14sn29} illustrates examples of the reflected MPCs at snapshots~15 and~29, overlaid on the captured Lidar point cloud. The measured and estimated VSs of the MPCs are represented by `*' and square, respectively. 
For better visualization, the figure omits those reflected MPCs interacting with the ceiling and the floor. It can be found that the side walls of the room contribute with dominant reflections to the channels. Furthermore, multi-bounce reflections are observed, e.g., VS~3 in the left-hand figure, which is postulated as a mirror image of the UE as a second-order reflection.
Different reflected MPCs are observed in different snapshots, which indicates that the channels are dynamic due to the movement of the UE. In addition, the consistent geometry relationship between the estimated VSs and the UE position validates the effectiveness of the presented MPC interaction mechanism identification framework. 
\begin{figure}[t]
	\centering
	{
		\begin{minipage}[tb]{0.35\textwidth}
			\centering
			\includegraphics[width=1\textwidth]{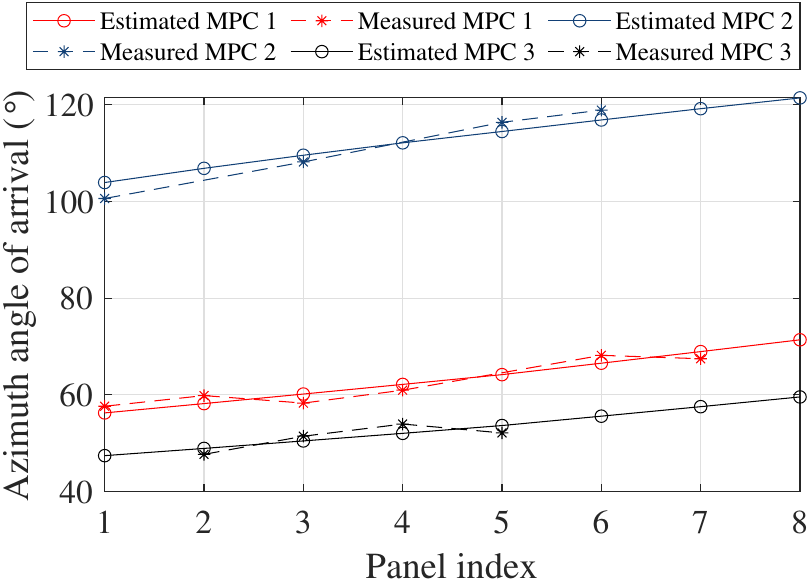}
		\end{minipage}
	}
	\caption{Example of measured and estimated AoAs of the reflected MPCs at snapshot~15.}
	\label{Fig_reflected_mpc_aoa}
\end{figure}
\begin{figure}[t]
\centerline{\includegraphics[width=0.35\textwidth]{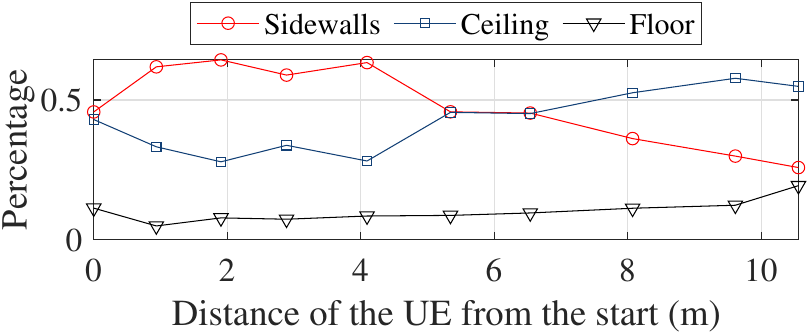}}
	\caption{Distribution of the reflected MPC v.s. UE moving distance.}
	\label{Fig_reflected_mpc_distribution}
\end{figure}  

Furthermore, the azimuth AoAs of the reflected MPCs at snapshot 15 are shown in Fig.~\ref{Fig_reflected_mpc_aoa}. For clarity, MPC~1, MPC~2, and MPC~3 refer to the MPCs associated with the VS~1, VS~2, and VS~3 in Fig.~\ref{Fig_example_reflected_mpc_sn14sn29}, respectively. The AoAs of `measured' MPCs are the results directly extracted from the measurements based on the SAGE algorithm, while the AoAs of `estimated' MPCs represent the results geometrically mapped from the estimated VSs. As expected, different panels are found to observe variant angles from the same VS, indicating the spherical wavefront of the MPC propagation. Moreover, the birth-and-death process of MPCs is observed in the measured channels, that is, some MPCs are only visible for part of the panels, and their lifetimes vary across the distributed panels due to the complex scattering environment. The good fit between the estimation results and the measurement data further validates the performance of the proposed framework.   

Finally, the spatial distribution of the reflected MPCs as the UE moves is shown in Fig.~\ref{Fig_reflected_mpc_distribution}. In the early stage of the UE movement, the highest percentage of reflected MPCs is observed from the sidewalls. When the UE moves towards the panels, the percentage of reflected MPCs from the ceiling is increasing. In the end, more than half of the reflected MPCs originate from the ceiling. This can be explained by the fact that with a shorter propagation distance, the possibility of MPCs interacting with the metal lamps that hang from the ceiling is greater. The dynamic distribution of the reflected MPCs indicates that the MPC gain and, consequently, the channels vary with the movement of the UE.   

\section{Conclusions}
In this letter, an indoor distributed MIMO channel measurement campaign has been carried out. The MPC characteristics of the channels have been investigated, focusing on the order and mechanisms of MPC interaction. 
The findings provide critical insights and statistical data that are valuable to develop an accurate and spatially consistent distributed MIMO channel model in future work.

\ifCLASSOPTIONcaptionsoff
  \newpage
\fi

\end{document}